# Income Inequality and Intergenerational Mobility in India

Anuradha Singh

Economics and Finance Department, BITS Pilani Campus

**Abstract**

Using three rounds of NSS datasets, the present paper attempts to understand the relationship between income inequality and intergenerational income mobility (IGIM) by segregating generations into social and income classes. The originality of the paper lies in assessing the IGIM using different approaches, which we expect to contribute to the existing literature. We conclude that the country has low-income mobility and high inequality which is no longer associated with a particular social class in India. Also, both may have a negative or positive relationship, hence needs to be studied at a regional level.

# Income Inequality and Intergenerational Mobility in India

## 1. Introduction

This paper examines the level of intergenerational income mobility (IGIM) among social classes in India and compares its relationship with income inequality. The IGIM analyses the growth of household units in terms of income over generations. If the growth of income is concentrated to certain sections of the society, there is immobility of income which further increases the level of inequality leading to more unequal distribution of resources. Therefore, low social mobility is both a cause and a consequence of rising inequalities and has adverse consequences for social cohesion and inclusive growth (Corak, 2016).

Looking at the case of India, the share of the top 10 per cent income group in the national income has increased and the share of the middle 40 per cent and bottom 50 per cent income groups has declined, as noted in Chancel and Piketty (2017). In addition, Ray (2014) estimated the relative IGIM to decline from 20 per cent in 1993 to 18 per cent in 2009. While in Scandinavia countries, children benefit from the parents' earnings, but parents' income or prestige in society does not determine the child's future income or job (Corak, 2013). In contrast to this in India, a recent study has concluded occupational immobility, implying that the occupation of the father largely determines the occupation of the child (Reddy, 2015).

Furthermore, despite the low economic mobility in the country, Hnatkovska et al. (2013) shows that the gap between the most disadvantaged social class (ST/SCs) and general (non-ST/SCs) in the country has narrowed. Also, the elasticity of wages for children in relation to the wages of their parent has declined from 88 per cent to 45 per cent for ST/SCs and from 76 to 58 per cent for non-ST/SCs. Therefore, ST/SC children are more likely to improve their relative position in income distribution than non-ST/SC children. However, Li et al. (2019) suggested that ST/SCs still have the least chance of escaping poverty and the highest probability of entering poverty. In light of the above contradictory findings, we expect that this study will contribute to the existing literature by incorporating income inequality with intergenerational income mobility across income groups of different social classes.

Although, there are some studies which talk about the two concepts separately, there is a paucity of studies that talk about both together. Further, due to the lack of appropriate data in the country for this study, we measure IGIM using two approaches: relative income mobility and absolute income mobility, where the first approach is related with how much of an individual's income is determined by his parent's income, the second is concerned with the ability of an individual to earn more than his parents at the same age.

## 2. Data and methodology

The data used in this paper are from the 43$^{rd}$ (1987-88), 61$^{st}$ (2004-05) and 68$^{th}$ (2011-12) rounds of the Employment and Unemployment survey (EUS) conducted by the National Sample Survey (NSS) of India. The EUS provides the primary source of data for various indicators of labour force at the state and national level. It follows a stratified multi-stage sample design and includes a sample of approximately 100,000 households covering almost all geographic regions of the country. This is the largest data collection in India since 1983 on almost every social and economic aspect at the individual and household level. However, the NSS does not collect information about parents if the individual is living separately from his family. Also, we concentrate on male subjects because married women in India live with their husbands or father-in-law and the survey does not provide information on their parents. Therefore, to estimate intergenerational income mobility (IGIM), we followed two approaches as previously described. In the first approach, we estimate the relative income mobility based on a sample of co-resident households and in the second approach, we measure the absolute mobility between two generations using independent samples. Further, using the 61$^{st}$ round, we measure the Gini coefficient on a monthly per capita expenditure (MPCE) basis.

2.1. Relative Income Mobility

In this approach we used the 68$^{th}$ round of NSS data and selected only those households where the working person and his father are living together. The criteria for selection of the working sample were households where the age of the son was between 16 and 45 and both father and son were not currently enrolled in any educational institution and informed about their wages. The proxy for income is the wages of individuals. This process yielded a sample of 10364 observations, which is the 'working sample' for co-resident households. To test whether sample selection is not biased, we compared the socioeconomic characteristics of the co-resident sample with sons who were living separately from their fathers (non-Co-resident sons), as shown in Table 1. In practice, non-co-resident sons are the households with only one adult male who is of working age. We found 48390 records for non-co-resident households. Their comparison does not show significant differences in terms of caste, rural-urban structure, education and consumption, proving that split decisions are random.

In addition, the descriptive data from the co-resident sample indicate that the average age of sons is 24 and father is 52. The mean years of education in the sons' generation is 9.93 years, while it is only 6.42 years in the father's generation. If we look at the occupation, the sons' generation is governed by skilled and semi-skilled occupations, while the father's generation is mostly in the farming occupation which is 39 per cent.

Table1: Summary statistics of co-resident sons and sons who are living on their own

|  | Co-resident | | Living on their own | |
| --- | --- | --- | --- | --- |
| Variable | Mean | SD | Mean | SD |
| Age | 25.91 | 6.12 | 35.83 | 6.54 |
| % Of Rural Pop. | 69.72 | … | 57.87 | … |
| % Of ST/SC | 27.24 | … | 35.83 | … |
| Years of Education | 9.94 | 3.41 | 8.84 | 4.46 |
| Log MPCE | 7.14 | 0.54 | 7.26 | 0.59 |

**Source:** Author's calculation

With respect to income, there is a significant difference in the age of father and son so the average income of father and son is also quite different. Since the main objective is to provide comparable estimates of income mobility in India across social classes, selection bias, if any, would affect all groups, therefore the inferences drawn will be robust.

Next, we classified the incomes of fathers and sons into four income groups according to Björklund and Jäntti (1997) distribution criteria. Here, (1) the poor are defined as those earning less than 50 per cent of the average income; (2) lower middle income, from 50 per cent of the average to the average, (3) upper-middle income, 1.0 to 1.5 times the average, (4) well to do, earning more than 150 per cent above average. The above classification was used to compute the cross-table of the father and child quintile groups, providing its probability distribution. It shows how the current generation performs among their peers as compared to their father's position among their peers.

2.2. Absolute Wage Income Mobility

In this approach we used MPCE as a proxy for income. We selected father and son data from two rounds which are 43 and 68 respectively. From the 43$^{rd}$ round (1987-88) we selected individuals who were between 35 and 40 years of age and also had a son between 10 and 15 years of age, so that in 1987-88 year a male child of 10 to 15 years of age would attain the age of 35 to 40 in 2011-12 year (68$^{th}$ round). In the case where the father had more than one son, only one son was selected at random. This selection leaves us with a working sample of 7928 observations from the eight states of the country. The eight states were selected based on high, medium, and low per capita GDP as according to constant 2011-12 prices by the Central Statistical Organization (CSO). The states selected are: Assam, Kerala, Maharashtra, Odisha, Rajasthan, Tamil Nadu, Uttar Pradesh and West Bengal. The age of 35 to 40 was selected for recording the income because at this age income of a person subjects to minimum life cycle bias.

The selection of samples from two independent samples is valid on the grounds that it involves comparison of the weighted average income of father and son on state level, therefore, it does not require data on pairs of fathers and of sons. Moreover, the segregation of the state-level data into social classes- STs, SCs and non-ST/SCs- enabled comparisons between these groups across states and within a state. Scheduled Tribes (STs) and Scheduled Castes (SCs) are among the most disadvantaged groups in India while non-ST/SCs are considered to be economically better off than ST/SCs. In this way it was possible to filter the records of fathers and sons from two independent samples, allowing us to estimate absolute income mobility at the state level, where both fathers and sons are of the same age and belong to the generations before (1987-88) and after (2011-12) liberalization and structural reforms respectively.

In addition, socio-economic characteristics of the data in Table2 show that except the state of Assam and Kerala, population of ST/SC and rural areas is same in both the samples of data. This is because the state of Assam saw an influx of migrants from neighbouring countries, while migration of people from Kerala was recorded. As per census records, migration of people is mostly within the state except in these two states.

Subsequently, in order to compare the real average income of father and son from the two rounds, the old year values of MPCE from the year 1987-88 were converted to the new year prices for 2011-12. For this, the state level Consumer Price Index for Agricultural Labors (CPIAL) used for rural areas and Consumer Price Index for Industrial Workers (CPI-IW) used for urban areas and then linking factors were used to equate the base year.

The following formula was used to convert the values of old year into new year prices:

**$Year\ 1987 - 88$ value in $2011 - 12$ prices**

$$= 1987 - 88\ Year\ Prices \times \frac{Index\ Number\ for\ 2011 - 12\ Year}{Index\ Number\ for\ 1987 - 88\ Year}$$

Next, the MPCE values were assigned weights to calculate weighted average income of STs, SCs and non-ST/SCs within each state. Then, we compared these values at the state level to record the difference in the mean income of people in the same social group between two generations. If the weighted average income is higher than that of the father generation (say 10 per cent or more), then upward mobility is suggested.

Table2: Summary statistics of father and son in the second approach

| State | Variable | Father Mean | Father SD | Son Mean | Son SD |
|---|---|---|---|---|---|
| Assam | Age | 38.22 | 1.82 | 37.44 | 1.94 |
|  | % Of rural population | 79.27 | … | 78.82 | … |
|  | % Of ST/SC | 23.5 | … | 32.1 | … |
|  | Years of Education | 6.33 | 6.91 | 9.81 | 4.12 |
|  | Log MPCE | 5.08 | 0.46 | 6.98 | 0.47 |
| Kerala | Age | 38.15 | 1.82 | 37.56 | 1.78 |
|  | % Of rural population | 75 | … | 75 | … |
|  | % Of ST/SC | 13.54 | … | 9.38 | … |
|  | Years of Education | 9.57 | 4.05 | 9.16 | 2.85 |
|  | Log MPCE | 6.7 | 0.9 | 7.46 | 0.57 |
| Maharashtra | Age | 37.87 | 2.09 | 37.49 | 2.05 |
|  | % Of rural population | 54.3 | … | 54.03 | … |
|  | % Of ST/SC | 23.7 | … | 21.4 | … |
|  | Years of Education | 7.11 | 4.5 | 8.96 | 3.81 |
|  | Log MPCE | 5.16 | 0.59 | 7.35 | 0.6 |
| Odisha | Age | 37.73 | 2.03 | 37.42 | 2 |
|  | % Of rural population | 72.99 | … | 72.99 | … |
|  | % Of ST/SC | 35.77 | … | 37.96 | … |
|  | Years of Education | 6.2 | 5.99 | 8.42 | 4.48 |
|  | Log MPCE | 6.19 | 0.716 | 6.85 | 0.52 |
| Rajasthan | Age | 37.88 | 2.12 | 37.52 | 2.14 |
|  | % Of rural population | 64.36 | … | 64.36 | … |
|  | % Of ST/SC | 29.5 | … | 27.7 | … |
|  | Years of Education | 5.13 | 4.91 | 7.64 | 5.04 |
|  | Log MPCE | 5.15 | 0.68 | 7.21 | 0.5 |
| Tamil Nadu | Age | 38.09 | 1.92 | 37.7 | 1.87 |
|  | % Of rural population | 52.8 | … | 52.8 | … |
|  | % Of ST/SC | 21.2 | … | 21.4 | … |
|  | Years of Education | 6.42 | 4.54 | 9.19 | 4.29 |
|  | Log MPCE | 5.1 | 0.71 | 7.33 | 0.55 |
| Uttar Pradesh | Age | 38 | 2.15 | 37.29 | 2.07 |
|  | % Of rural population | 67.5 | … | 66.7 | … |
|  | % Of ST/SC | 23.16 | … | 20.47 | … |
|  | Years of Education | 5.84 | 5.82 | 7.56 | 5.22 |
|  | Log MPCE | 6.32 | 0.802 | 6.94 | 0.531 |
| West Bengal | Age | 37.67 | 2.01 | 37.62 | 1.9 |
|  | % Of rural population | 62.55 | … | 62.55 | … |
|  | % Of ST/SC | 28.5 | … | 29 | … |
|  | Years of Education | 5.88 | 4.55 | 8.66 | 4.96 |
|  | Log MPCE | 5.07 | 4.55 | 7.16 | 0.58 |

**Source:** Author's calculation

## 2.3. Income Inequality

Using the 61$^{st}$ round (2004-05) of the EUS from NSS data, the Gini coefficient was calculated for STs, SCs and non-ST/SCs in eight states of the country. Here also MPCE was used as a proxy for income to calculate income inequality. The Gini coefficient ranges from zero to one where zero denotes perfect equality i.e., each income quintile has the same income and one denotes perfect inequality where the top income quintile generates all incomes. It was calculated using the following formula:

$$\text{Gini Coefficient} = \frac{Area\ between\ Perfect\ Equality\ Lorenz\ and\ Actual\ Lorenz}{Area\ under\ Perfect\ Equality\ Lorenz}$$

where,

Area under Perfect Equality Lorenz = 1/2 (Side × Side)

$$\text{Area under the Actual Lorenz Curve} = \text{Bar Width}^1 \times \text{Bar Height}^2$$

The average Gini Coefficient for the states was verified by the estimates of Planning Commission, Government of India. Below 0.30 it is assumed that there is low-income inequality and above there is high-income inequality.

## 3. Relative Intergenerational Income Mobility

Table 3 Panel A. reports the highest percentage of poor as well as well-to-do fathers and sons among STs. Also, the highest percentage of well-to-do sons among STs coincides with the proportion of well-to-do sons among non-ST/SCs, which is 11.6 per cent. Overall, the proportion of lower-middle class is the highest among both father and son generations. Furthermore, the mobility pattern in Panel B shows the probability of a son ending up in a specific income class, conditional on his father's income class. It shows that the probability of being poor if the father is poor is 73.3 per cent while the probability of being rich if the father is rich is 38.7 per cent. Here, within each income class it is showing immobility for all social groups. STs are most likely to be poor from a poor father, which is 81.9 per cent, while non-ST/SCs are more likely to be rich from a rich father, which is 42.3 per cent. In addition, the chances for ST/SCs to improve from lower-middle background to an upper-middle are less than for non-ST/SCs. If only income distribution is taken into account, a son in the lower middle class is more likely to be in the lower middle again. Further, higher middle class are more likely to end up in the low or higher middle than in the well-to-do class. Also, sons from well-to-do backgrounds are more likely to be either in the lower-

---

[1] The Bar Width is estimated using the cumulative percentage population difference
[2] The Bar Height is measured as the average of the cumulative percentage income

Table 3: Mobility Matrices

| Father's income class | Son's income class | | | |
|---|---|---|---|---|
| | Poor | Lower-middle | Higher-middle | Well-to-do |
| Panel A. Unconditional bivariate probabilities | | | | |
| **All Social Groups** | | | | |
| Poor | 0.220 | 0.067 | 0.009 | 0.004 |
| Lower-middle | 0.093 | 0.273 | 0.023 | 0.009 |
| Higher-middle | 0.015 | 0.042 | 0.030 | 0.004 |
| Well-to-do | 0.037 | 0.065 | 0.028 | 0.082 |
| **Scheduled Tribes (STs)** | | | | |
| Poor | 0.281 | 0.055 | 0.005 | 0.002 |
| Lower-middle | 0.072 | 0.214 | 0.007 | 0.005 |
| Higher-middle | 0.012 | 0.043 | 0.026 | 0.002 |
| Well-to-do | 0.064 | 0.076 | 0.029 | 0.107 |
| **Scheduled Castes (SCs)** | | | | |
| Poor | 0.229 | 0.071 | 0.008 | - |
| Lower-middle | 0.091 | 0.330 | 0.015 | 0.008 |
| Higher-middle | 0.018 | 0.043 | 0.036 | 0.003 |
| Well-to-do | 0.036 | 0.052 | 0.020 | 0.039 |
| **Others** | | | | |
| Poor | 0.205 | 0.067 | 0.01 | 0.006 |
| Lower-middle | 0.097 | 0.261 | 0.029 | 0.01 |
| Higher-middle | 0.015 | 0.041 | 0.028 | 0.004 |
| Well-to-do | 0.032 | 0.069 | 0.032 | 0.096 |
| Panel B. Son's probability conditional on father's income | | | | |
| **All Social Groups** | | | | |
| Poor | 0.733 | 0.223 | 0.030 | 0.013 |
| Lower-middle | 0.234 | 0.686 | 0.058 | 0.023 |
| Higher-middle | 0.166 | 0.466 | 0.333 | 0.044 |
| Well-to-do | 0.175 | 0.307 | 0.132 | 0.387 |
| **Scheduled Tribes (STs)** | | | | |
| Poor | 0.819 | 0.16 | 0.015 | 0.006 |
| Lower-middle | 0.242 | 0.718 | 0.023 | 0.017 |
| Higher-middle | 0.145 | 0.518 | 0.313 | 0.02 |
| Well-to-do | 0.232 | 0.275 | 0.105 | 0.388 |
| **Scheduled Castes (SCs)** | | | | |
| Poor | 0.744 | 0.231 | 0.026 | - |
| Lower-middle | 0.204 | 0.742 | 0.034 | 0.018 |
| Higher-middle | 0.18 | 0.43 | 0.36 | 0.03 |
| Well-to-do | 0.245 | 0.354 | 0.136 | 0.265 |
| **Others** | | | | |
| Poor | 0.709 | 0.232 | 0.035 | 0.021 |
| Lower-middle | 0.244 | 0.657 | 0.073 | 0.025 |
| Higher-middle | 0.172 | 0.471 | 0.322 | 0.046 |
| Well-to-do | 0.141 | 0.304 | 0.141 | 0.423 |

**Source:** Author's calculation

middle or well-to-do class. Overall, it can be suggested that downward mobility is more probable than upward mobility in co-resident households.

**4. Income Inequality and Absolute Intergenerational Mobility**

Figure 1 shows the improvement in the average income of the son over the father's generation in the case of Tamil Nadu, Uttar Pradesh and West Bengal for both ST/SCs and non-ST/SCs, while the level of inequality in Table 4 is also very high in these states. Of all the states, Tamil Nadu shows the highest margin of improvement in income for all social classes. The state of Rajasthan records immobility for non-ST/SCs and high inequality for them. In addition, income mobility for ST/SCs in the state was observed with relatively low inequality for them, indicating a positive relationship between income inequality and income mobility. Furthermore, Maharashtra is in contrast to Rajasthan in that the average income of sons has improved for non-ST/SCs and not for ST/SCs. With regards to inequality, Maharashtra records a high level of inequality for all social classes while Rajasthan has a low overall inequality. Further, the state of Assam does not show any improvement for any social class which may be due to the considerable change in the population size of the state due to immigration from the bordering countries. Therefore, it is not considered a valid case to compare the absolute income of the state between the two generations. Interestingly, this state also has the lowest level of income inequality among all social classes as compared to other states of the country. Kerala shows high income inequality and immobility for STs and non-ST/SCs while income mobility and low inequality for SCs. Odisha shows high income inequality for all social classes and immobility for SCs and mobility for ST and non-ST/SCs. Overall, it is concluded that, except the state of Tamil Nadu, all other seven states do not show significant improvement even with absolute income between two generations. Thus, there is low-income mobility and high inequality. Also, the improvement or reduction in the absolute income levels for ST/SC in different states of the country is not much different from that for non-ST/SCs.

Figure1: Weighted Average Income of Parent and Children or absolute income mobility

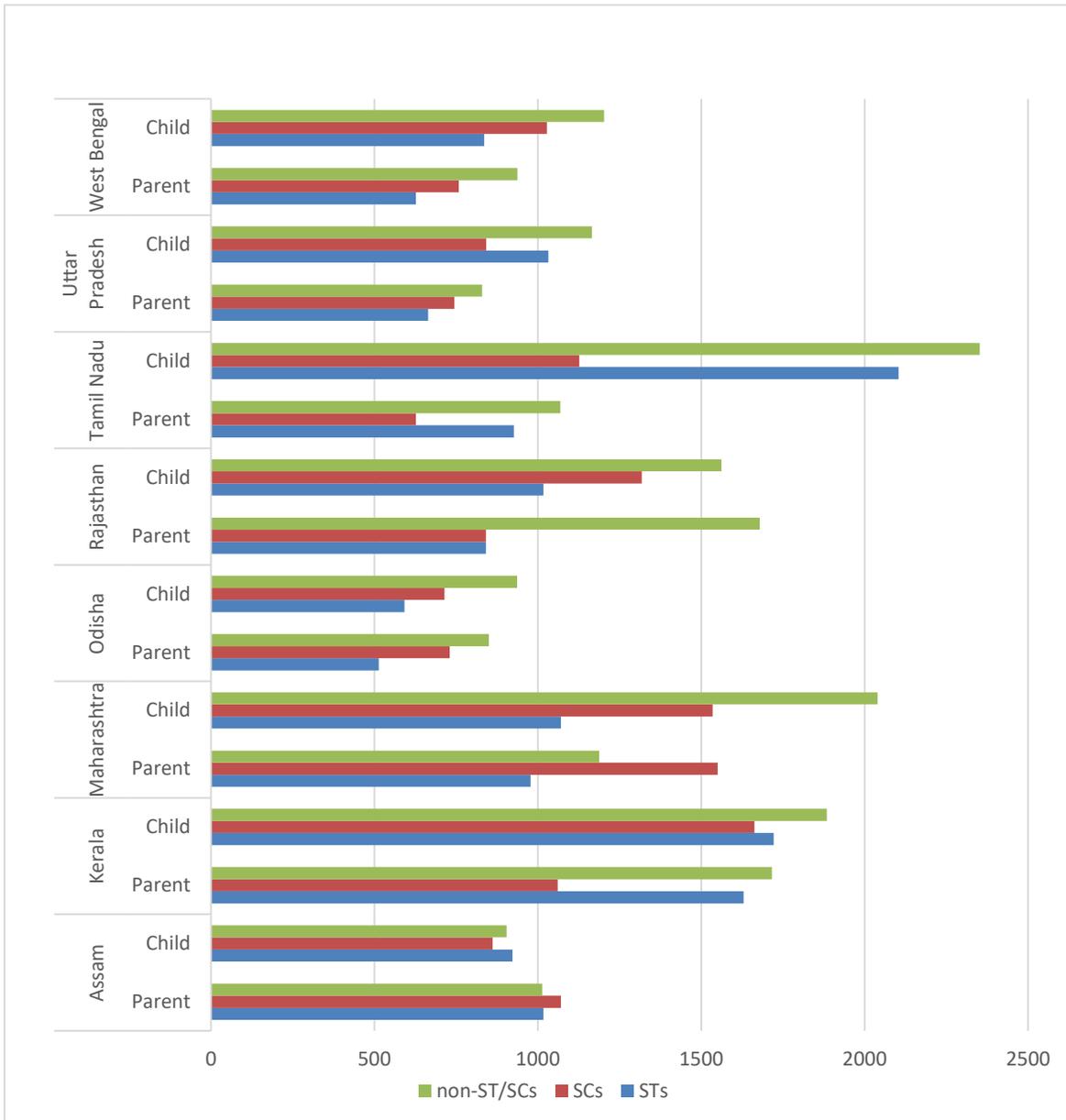

**Source:** Author's calculation

Table 4: Gini Coefficients

| States | Overall | STs | SCs | Non-ST/SCs |
|---|---|---|---|---|
| Assam | 0.27 | 0.24 | 0.26 | 0.31 |
| Kerala | 0.36 | 0.44 | 0.30 | 0.35 |
| Maharashtra | 0.36 | 0.34 | 0.36 | 0.39 |
| Odisha | 0.36 | 0.32 | 0.41 | 0.34 |
| Rajasthan | 0.29 | 0.28 | 0.28 | 0.32 |
| Tamil Nadu | 0.37 | 0.40 | 0.33 | 0.38 |
| Uttar Pradesh | 0.33 | 0.36 | 0.29 | 0.34 |
| West Bengal | 0.33 | 0.33 | 0.31 | 0.36 |
| India | 0.34 | 0.34 | 0.33 | 0.37 |

**Source:** Author's calculation

## 5. Concluding Remarks

In this paper we have investigated intergenerational income mobility and inequality across social classes in India by using the 43$^{rd}$, 61$^{st}$, and 68$^{th}$ rounds of NSS data on EUS. Our results indicate that in India there is low intergenerational income mobility and high-income inequality, which is also true for all social classes in the country. Further, relative income mobility shows more immobility for ST/SCs than non-ST/SCs. Also, the absolute income mobility between the generations from pre-reform period and one from after twenty years of reforms, does not show much improvement in the real average income levels of two generations from different states. In addition, it shows overall similar improvement and reduction for all social classes in the country. However, it is true that few regions certainly lag behind and still needs to focus on reducing the gap between social classes. Therefore, high income inequality and low intergenerational mobility coupled with high economic growth point towards unequal growth in the country.

With regards to the regional level, the assumption that higher unequal regions always have lower mobility does not hold true in the case of India. As, there are few states in the country which are having high income inequality with income mobility. Thus, this finding emphasizes on the conclusion that inequality and intergenerational mobility is a local phenomenon which needs to be studied at a regional level. In addition, greater mobility in the most unequal regions can be associated with inequality due to the rapid expansion of upper quartiles, which needs to be examined further and also whether income mobility with lower inequality is associated with inclusive growth and development of the region.

References:

Björklund, A., & Jäntti, M. (1997). Intergenerational Income Mobility in Sweden Compared to the United States. *American Economic Review*, *87*(5), 1009–1018. https://doi.org/10.2307/2951338

Chancel, L., & Piketty, T. (2017). *Indian income inequality , 1922-2015 : From British Raj to Billionaire Raj ? Lucas Chancel* (2017/11, Issue July).

Corak, M. (2013). Income Inequality, Equality of Opportunityand Intergenerational Mobility. *Journal of Economic Perspectives*, *27*(3), 79–102. https://doi.org/10.1257/jep.27.3.79

Corak, M. (2016). Inequality from Generation to Generation: The United States in Comparison. *IZA Discussion Paper*, *9929*.

Hnatkovska, V., Lahiri, A., & Paul, S. B. (2013). Breaking the caste barrier: intergenerational mobility in India. *The Journal of Human Resources*, *48*(2), 435–473. https://doi.org/10.1353/jhr.2013.0012

Li, H., Millimet, D. L., & Roychowdhury, P. (2019). Measuring Economic Mobility in India Using Noisy Data : A Partial Identification Approach. *IZA Discussion Paper No. 12505*, *12505*. https://ssrn.com/abstract=3435380

Ray, J. (2014). *Intergenerational Mobility in Educational and Occupation: A Study of Social Classes in India* [The University of Burdwan]. https://doi.org/http://hdl.handle.net/10603/201955

Reddy, A. B. (2015). Changes in Intergenerational Occupational Mobility in India: Evidence from National Sample Surveys, 1983-2012. *World Development*, *76*, 329–343. https://doi.org/10.1016/j.worlddev.2015.07.012